\documentclass{jfm}
\usepackage{multirow,bigdelim}
\usepackage{graphicx}
\usepackage{natbib}
\usepackage{psfrag}
\usepackage{color}
\usepackage{enumerate}
\usepackage{rotating,pdflscape}
\ifCUPmtlplainloaded \else
  \checkfont{eurm10}
  \iffontfound
    \IfFileExists{upmath.sty}
      {\typeout{^^JFound AMS Euler Roman fonts on the system,
                   using the 'upmath' package.^^J}%
       \usepackage{upmath}}
      {\typeout{^^JFound AMS Euler Roman fonts on the system, but you
                   dont seem to have the}%
       \typeout{'upmath' package installed. JFM.cls can take advantage
                 of these fonts,^^Jif you use 'upmath' package.^^J}%
       \providecommand\upi{\pi}%
      }
  \else
    \providecommand\upi{\pi}%
  \fi
\fi

\ifCUPmtlplainloaded \else
  \checkfont{msam10}
  \iffontfound
    \IfFileExists{amssymb.sty}
      {\typeout{^^JFound AMS Symbol fonts on the system, using the
                'amssymb' package.^^J}%
       \usepackage{amssymb}%

      }{}
  \fi
\fi

\ifCUPmtlplainloaded \else
  \IfFileExists{amsbsy.sty}
    {\typeout{^^JFound the 'amsbsy' package on the system, using it.^^J}%
     \usepackage{amsbsy}}
    {}
\fi

\newcommand\Rey{\mbox{\textit{Re}}}  

\newsavebox{\astrutbox}
\sbox{\astrutbox}{\rule[-5pt]{0pt}{20pt}}

\newcommand\eg{e.g.\ }

\title[A fast method for characterising hydraulic roughness]{A fast direct numerical simulation method for characterising hydraulic roughness}

\author[D. Chung, L. Chan, M. MacDonald, N. Hutchins and A. Ooi]%
{D. Chung%
  \thanks{Email address for correspondence: daniel.chung@unimelb.edu.au},\ns
L. Chan,\ns M. MacDonald,\ns N. Hutchins and A. Ooi}

\affiliation{Department of Mechanical Engineering, University of Melbourne,
Victoria 3010, Australia}

\pubyear{2010}
\volume{650}
\pagerange{119--126}
\date{?; revised ?; accepted ?. - To be entered by editorial office}
\begin{document}

\maketitle

\begin{abstract}
We describe a fast direct numerical simulation (DNS) method that
promises to directly characterise the hydraulic roughness of any
given rough surface, from the hydraulically smooth to the fully rough
regime.
The method circumvents the unfavourable computational cost associated with
simulating high-Reynolds-number flows by employing minimal-span
channels \citep{Jimenez+Moin.1991}.
Proof-of-concept simulations
demonstrate that flows in minimal-span channels
are sufficient for capturing the downward velocity shift, that is,
the Hama roughness function, predicted by
flows in full-span channels.
We consider two sets of simulations, first with modelled roughness imposed by body forces, and second with explicit roughness described by roughness-conforming grids.
Owing to the minimal cost, we are able
to conduct DNSs with increasing
roughness Reynolds numbers while maintaining a
fixed blockage ratio, as is typical in full-scale applications.
The present method promises a practical, fast and accurate
tool for characterising hydraulic resistance directly from
profilometry data of rough surfaces.
\end{abstract}

\begin{keywords}
Authors should not enter keywords on the manuscript, as these must be chosen by the author during the online submission process and will then be added during the typesetting process (see http://journals.cambridge.org/data/\linebreak[3]relatedlink/jfm-\linebreak[3]keywords.pdf for the full list)
\end{keywords}

\section{Introduction}\label{sec:introduction}

Scientists have, for years, been documenting the
relationship between surface roughness and its hydraulic resistance,
the former pertaining to geometry of the surface while the latter is dictated by the dynamics of flow over the roughness
\cite[see reviews by][]{Jimenez.2004,Flack+Schultz.2010}. The cataloguing process
can be a never-ending exercise because the characteristics of each rough surface are unique.
In order to make predictions on full-scale operations, it is necessary to establish the equivalent sand-grain roughness $k_s$ of a given surface, which relates the drag increment of the given surface to an equivalent surface composed of monodisperse sand grains. Once $k_s$ has been determined, it is possible to predict the drag penalty at application Reynolds numbers using either the Moody chart \citep{M1944}, for pipe and channel flows, or developments of this for zero-pressure-gradient boundary layers \citep{PS55,G58}. Generally speaking, the approach to date has been to first identify a particular rough surface of scientific or engineering interest, and then to characterise its hydraulic resistance through well-controlled laboratory experiments. By exposing the rough surface to an increasing range of flow speeds, until such point beyond which the resistance coefficient $C_f$ becomes constant, referred to as the `fully rough' asymptote, it is possible to ascertain $k_s$. A convenient dimensionless group here is the equivalent roughness Reynolds number $k_s^+ \equiv k_s U_\tau /\nu$, where the friction velocity
$U_\tau \equiv \sqrt{\tau_0/\rho} \equiv U_\infty\sqrt{C_f/2}$; $\tau_0$
is the wall drag force per plan area; $\rho$ is the density; $U_\infty$ is the freestream velocity or centreline velocity for internal geometries; and $\nu$ is the kinematic viscosity. For accurate predictions, it is not sufficient to merely establish $k_s$, rather the increment of $C_f$ caused by surface roughness must be mapped with respect to $k_s^+$ at all conditions from the dynamically smooth up to the fully rough regimes, which generally covers the approximate range $5 \lesssim k_s^+  \lesssim 100$. Over the last century or so, this painstaking and time-consuming procedure has been repeated for many surfaces of interest and as a result, there is now a rich database of
roughness amassed in the published literature over time.

One problem here is that there is no widely applicable function that relates $k_s$, a dynamic parameter, to readily observable or measurable geometric properties of the surface, such as the root-mean-square roughness height $k_{rms}$ or average roughness height $k_a$. Certain classes of surfaces, say sand-grain roughness, may exhibit an approximate proportionality between $k_s$ and some physical surface length scale
, but as a general rule this proportionality will not hold across different surface topologies
. There are numerous attempts in the literature to formulate more complicated functions, often involving some measure of mean surface height such as $k_{rms}$ or $k_a$ and other properties relating to the shape or arrangement of roughness elements such as skewness, effective slope, solidity and so on \citep[see review by][]{Flack+Schultz.2010}. Though many of these parameters have some success in describing the particular class of surfaces for which they were formulated (e.g.\ painted or sanded surfaces), none are widely applicable across the almost limitless range of surface topologies that are encountered in engineering and meteorological applications.

The present method to directly evaluate $k_s$ represents
a paradigm shift
from such parameterisations that are based on a handful
of geometrical factors.
Recognising that all roughness geometries are unique, and that a one-size-fits-all formulaic solution is proving elusive, we have sought an approach to minimise the expense involved in experimentally determining $k_s$.
Specifically, the present method relates raw profilometry data, which
is described by many degrees of freedom, directly to $k_s$.
In contrast, previous indirect approaches first reduce
raw profilometry data to a limited subset of geometrical factors such
as $k_{rms}$ and skewness before relating these factors to $k_s$.
The inevitable loss of information in reducing raw profilometry data
to geometrical factors is avoided by the present method
and, in this sense, the present method represents
a direct evaluation of the roughness function.


The approach relies on the direct computation
of hydraulic resistance by direct numerical simulation (DNS), which we
presently show can be made substantially cheaper, and therefore faster,
than previously thought.
We circumvent the otherwise-prohibitive cost of conventional DNS
by employing minimal-span channels \citep{Jimenez+Moin.1991,Hwang.2013},
the rationale of which is discussed in the following.
Normally, a straightforward and direct computation of roughness drag using DNS
employing full-span channels is extremely expensive as it entails
simultaneously capturing
both the bulk flow, which scales with the half-channel height, $h$,
and the near-wall flow around the roughness elements, which scales
with a characteristic roughness height, $k$.
Given a rough surface of fixed blockage ratio
$k_s/h \lesssim 1/40$ \citep*{Jimenez.2004,Flack+Schultz+Shapiro.2005}, a complete characterisation of
hydraulic resistance requires parametric simulations that sweep
through the equivalent roughness Reynolds numbers, $k_s^+  \approx
5$ to $100$, corresponding to the hydraulically smooth
and the fully rough regimes, respectively. For the blockage ratio, $h/k_s = 40$, this means performing parametric
simulations at the friction Reynolds numbers, $\textit{Re}_\tau
\equiv h U_\tau / \nu = (h/k_s) (k_s U_\tau/\nu) \approx 200$ to $4000$,
which are currently unfeasible.
Recall that the cost of DNS, counting the number of spatial and temporal degrees of freedom,
scales unfavourably as $\textit{Re}_\tau^3$ \citep[][\S\,9.1.2]{Pope.2000}. For grids conforming to the surface of the roughness elements, this cost is further exacerbated by the need for increased mesh density, and reduced time steps.

However, the extreme cost associated with conventional DNS employing
full-span channels seems unnecessary.
The quantity of interest from an engineering point of view
is the retardation in the mean flow over the roughness relative to
the smooth-wall flow.
This relative flow retardation or downward velocity shift,
$\Delta U$, occurs mostly in the vicinity of the roughness layer,
but holds constant above a few roughness heights, well
into the log layer (if it exists) and the wake region,
a behaviour described by
Townsend's outer-layer similarity hypothesis \citep{Townsend.1976}.
This suggests that a simulation of only the near-wall region
and its interaction with the roughness geometry is required in order
to extract $\Delta U^+ \equiv \Delta U/U_\tau$,
which is known as the (Hama) roughness function.
This distillation of the problem is consistent with
the observation that $\Delta U^+$ does not depend on the bulk
flow but only on $k_s^+$ and other details of the roughness geometry.

A framework for simulating only the near-wall dynamics is the minimal channel.
The concept is first described by \cite{Jimenez+Moin.1991} and is
currently receiving renewed attention
in various contexts of understanding wall-bounded turbulence
\citep{Flores+Jimenez.2010,Hwang.2013,Lozano-Duran+Jimenez.2014}.
Presently, we exploit this framework for
measuring $\Delta U^+$ by fully resolving the
near-wall Navier--Stokes dynamics and its interaction with
the roughness geometry.
The prohibitive cost of conventional DNS is
alleviated by use of these minimal-span channels,
which are designed to preclude the bulk flow that scales with $h$.
Without the bulk flow, the cost of DNS with roughness now only
scales as $k_s^{+3}$, which is quite feasible for the engineering
task at $k_s^+ \approx 5$ to $100$.
In principle, the computational cost is potentially $(h/k_s)^3$
times less than that
of a conventional DNS in a full-span channel.

In the remainder, we demonstrate the efficacy of this approach and
develop guidelines for its use. Beginning first with the parametric forcing model of
\cite{Busse+Sandham.2012}, we carefully confirm that the minimal channels return the same estimate for the roughness function as that given by full-span simulations (\S~\ref{sec:modelledroughness}). An important subtlety here lies in the choice of the sminimal spanwise unit, which we demonstrate is related not only to the usual span of the near-wall cycle (approximately 100 viscous wall units), but also to some physical roughness height $k$ and, presumably, the roughness texture under investigation. Having established the accuracy with the simple forcing model, we then test this procedure with the more realistic test case of a three-dimensional roughness, explicitly described by grids conforming to the rough surface, and compare results with full-domain simulations (\S~\ref{sec:griddedroughness}).

\section{Direct numerical simulations}\label{sec:simulations}

\begin{figure}
\centerline{\includegraphics{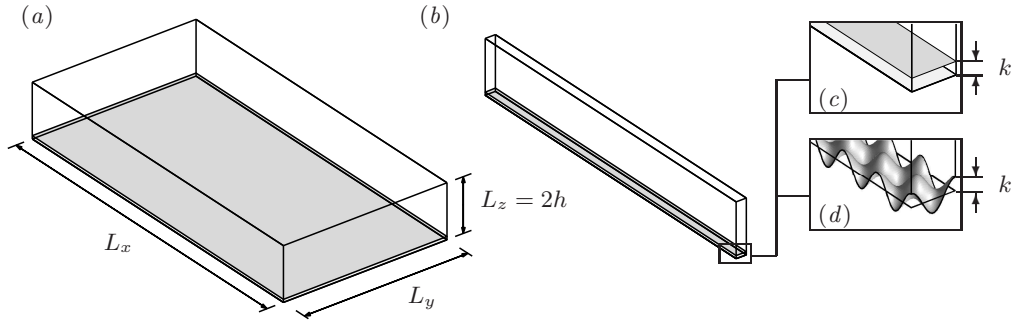}}
      \caption{Computational domain of (\mbox{\textit{a}}) the full channel and \mbox{(\textit{b}}) the minimal channel.
      The magnified view of the channel surface illustrates the rough wall represented by (\mbox{\textit{c}}) the
      roughness-forcing region, $0 < z < k$, or (\mbox{\textit{d}}) the explicit sinusoidal roughness with
      semi-amplitude $k$ and mean at $z = 0$. Both top and bottom channel walls are rough.}
      \label{fig:comp_dom}
\end{figure}

We present two sets of DNSs (table \ref{tab:cases}).
In Set 1, roughness is modelled by body forces
(figure \ref{fig:comp_dom}\mbox{\textit{c}})
in a fourth-order staggered-grid code as described by
\cite{Morinishi+coauthors.1998}. The code is written by
the first author and has been used in other studies \citep[\eg][]{Chung+Monty+Ooi.2014}.
In Set 2, roughness is explicitly represented by a roughness-conforming grid
(figure \ref{fig:comp_dom}\mbox{\textit{d}})
in the code CDP described by
\cite{Mahesh+Constantinescu+Moin.2004}.
The purpose of the Set-1 simulations with modelled roughness
is to test the present minimal-channel method for a hypothetical
roughness in the absence of any imposed wall-parallel length scales.
Such a configuration, which we shall call homogeneous roughness,
is ideal for investigating how the characteristic roughness height (as opposed to
wall-parallel length scales) sets the minimum allowable span $L_y$ of the channel
such that $\Delta U^+$ is still accurately captured.
The purpose of the Set-2 simulations with grids conforming to the rough surface
is to demonstrate that the present minimal-channel method works as
expected for a specific roughness geometry, namely the
sinusoidal `egg-carton' roughness that has previously been characterised (in a pipe) by
\cite{Chan+coauthors.2015}.

In both sets of simulations, we solve the following Navier--Stokes equations of motion:
\begin{subeqnarray}\label{eq:ns}
\gdef\thesubequation{\theequation \mbox{\textit{a}},\mbox{\textit{b}}}
\frac{\partial u_i}{\partial t}
  + \frac{\partial (u_j u_i)}{\partial x_j}
  = - \frac{1}{\rho}\frac{\partial p}{\partial x_i}
  + \nu \frac{\partial^2 u_i}{\partial x_j^2}
  + f(t)\,\delta_{i1}, \quad
\frac{\partial u_j}{\partial x_j} = 0,
\end{subeqnarray}
\returnthesubequation
where $u_i$ is the velocity; $t$ is time; $x_j$ is the spatial coordinate;
$p$ is the pressure; and $f(t)$ is the spatially uniform,
time-varying, mean pressure gradient that drives the flow at constant mass flux.
The streamwise, spanwise and wall-normal directions are referred to as either $x_1$, $x_2$ and $x_3$ or $x$, $y$ and $z$ respectively.
Periodic boundary conditions are imposed in the streamwise and spanwise
directions with the respective domain sizes, $L_x$ and $L_y$
(figure \ref{fig:comp_dom}\mbox{\textit{a}}, \mbox{\textit{b}}).
In the reference full-span simulations, $L_y$ measures $\pi h$
for Set 1 and $2 \pi h$ for Set 2 (at several selected $\Rey_\tau$),
while in the minimal-span simulations, $L_y$ is constrained by
the consideration of three issues:
\begin{enumerate}[(1) ]
\item the span of the near-wall cycle (assumed to be approximately $100\,\nu/U_\tau$),
\item the span of the roughness elements, $\lambda_y$ (absent for Set-1 simulations), and
\item the physical height of the roughness elements, $k$.
\end{enumerate}
Regarding point (1), previous studies
\citep{Jimenez+Moin.1991,Hwang.2013} have shown that $L_y^+ > 100$
or else the near-wall cycle, which is the signature of wall turbulence,
cannot be properly captured.
Regarding point (2), the span of the channel must be wide enough to
accommodate the widest scale of the roughness elements
in order to properly represent the nature of the roughness in question. For the Set-1 simulations, which have a homogeneous forcing, this point is not an issue.
Regarding point (3), it turns out that the height of the roughness elements
also plays a role in setting the minimum $L_y$, an issue that
will be discussed in \S\,\ref{sec:modelledroughness}.
The chosen resolutions (table \ref{tab:cases}) are comparable to other channel-flow
DNSs \citep[\eg][]{Moser+Kim+Mansour.1999,Bernardini+Pirozzoli+Orlandi.2014}
and the streamwise domain lengths are long enough to accommodate
several near-wall streaks, which are approximately $1000\,\nu/U_\tau$.
Although the present study focusses on
minimising the spanwise dimension of the channel, which is the
critical dimension in setting the dynamics of the log layer
\citep{Flores+Jimenez.2010},
one presumes that the streamwise dimension will also be
subject to constraints.
For a smooth wall, \cite{Chin+coauthors.2010} show
that the mean velocity profile is slightly overestimated
when $L_x^+ \lesssim 1000$ (at $\Rey_\tau \approx 180$ for pipe flow),
suggesting that the near-wall streaks need to be properly captured.
It is also clear that $L_x$ should be greater than $\lambda_x$, where $\lambda_x$ is the
largest characteristic streamwise scale of the roughness elements.
The characteristic roughness sizes $k$ are selected so that they are
a fixed and small fraction of the half-channel height $h$.
The roughness Reynolds number $k^+$ is increased by increasing the overall friction Reynolds number $\Rey_\tau$,
which is the typical way roughness is encountered
in practice (experimentally, and in practice, the roughness Reynolds number is usually increased by
increasing the flow speed).

\begin{table}
\newcommand{\s}{\hphantom{0}}
  \begin{center}
\def~{\hphantom{0}}
  \begin{tabular}{cccccccccccccc}

  & {Roughness}
             & Span    & $\Rey_\tau$
                                 & $h/k$
                                        & $L_x^+$
                                                 & $L_y^+$ & $N_x$ & $N_y$  & $N_z$   & $\Delta x^+$
                                                                                              & $\Delta y^+$
                                                                                                      & $\Delta z_w^+$
                                                                                                               & $\Delta z_c^+$
                                                                                                                         \\[3pt]
\ldelim\{{13}{8mm}[Set 1] & Modelled & Minimal & $\s180$ & $40$ & $3707$ & $\s116$ & $384$ & $\s24$ & $\s192$ & $9.7$ & $4.8$ & $0.02$ & $\s2.9$ \\[3pt]
 & Modelled & Minimal & $\s395$ & $40$ & $3707$ & $\s116$ & $384$ & $\s24$ & $\s192$ & $9.7$ & $4.8$ & $0.05$ & $\s6.5$ \\[3pt]
 & Modelled & Full    & $\s590$ & $40$ & $3707$ & $1854$  & $384$ & $384$  & $\s256$ & $9.7$ & $4.8$ & $0.04$ & $\s7.2$ \\
 & Modelled & Minimal & $\s590$ & $40$ & $3707$ & $\s116$ & $384$ & $\s24$ & $\s256$ & $9.7$ & $4.8$ & $0.04$ & $\s7.2$ \\[3pt]
 & Modelled & Full    & $\s950$ & $40$ & $5969$ & $2985$  & $640$ & $640$  & $\s384$ & $9.3$ & $4.7$ & $0.03$ & $\s7.8$ \\
 & Modelled & Minimal & $\s950$ & $40$ & $3581$ & $\s112$ & $384$ & $\s24$ & $\s384$ & $9.3$ & $4.7$ & $0.03$ & $\s7.8$ \\[3pt]
 & Modelled & Minimal & $2000$  & $40$ & $3707$ & $\s116$ & $384$ & $\s24$ & $\s768$ & $9.7$ & $4.8$ & $0.02$ & $\s8.2$ \\
 & Modelled & Minimal & $2000$  & $40$ & $3707$ & $\s232$ & $384$ & $\s48$ & $\s768$ & $9.7$ & $4.8$ & $0.02$ & $\s8.2$ \\
 & Modelled & Minimal & $2000$  & $40$ & $3707$ & $\s463$ & $384$ & $\s96$ & $\s768$ & $9.7$ & $4.8$ & $0.02$ & $\s8.2$ \\[3pt]
 & Modelled & Minimal & $4000$  & $40$ & $3707$ & $\s116$ & $384$ & $\s24$ & $1024$  & $9.7$ & $4.8$ & $0.02$ & $12.3$  \\
 & Modelled & Minimal & $4000$  & $40$ & $3707$ & $\s232$ & $384$ & $\s48$ & $1024$  & $9.7$ & $4.8$ & $0.02$ & $12.3$  \\
 & Modelled & Minimal & $4000$  & $40$ & $3707$ & $\s463$ & $384$ & $\s96$ & $1024$  & $9.7$ & $4.8$ & $0.02$ & $12.3$  \\[3pt]
 {} & {} & {} & {} & {} & {} & {} & {} & {} & {} & {} & {} & {} & {} \\
\ldelim\{{6}{8mm}[Set 2] & Explicit & Full    & $\s180$ & $18$ & $2262$ & $1131$  & $510$ & $254$  & $\s156$ & $4.4$ & $4.5$ & $0.13$ & $\s4.5$ \\
 & Explicit & Minimal & $\s180$ & $18$ & $4029$ & $\s141$ & $910$ & $\s32$ & $\s156$ & $4.4$ & $4.4$ & $0.13$ & $\s4.4$ \\[3pt]
 & Explicit & Minimal & $\s360$ & $18$ & $4100$ & $\s141$ & $924$ & $\s28$ & $\s314$ & $4.5$ & $5.0$ & $0.13$ & $\s5.0$ \\[3pt]
 & Explicit & Minimal & $\s540$ & $18$ & $4029$ & $\s212$ & $755$ & $\s42$ & $\s470$ & $5.2$ & $5.0$ & $0.13$ & $\s5.2$ \\[3pt]
 & Explicit & Minimal & $1080$  & $18$ & $4241$ & $\s424$ & $869$ & $\s84$ & $\s940$ & $4.8$ & $5.0$ & $0.13$ & $\s5.0$ \\
  \end{tabular}
  \caption{Simulation cases listed with nominal $\Rey_\tau$.
  Each of these 17 cases is run using both smooth and rough walls,
  that is, a total of 34 separate simulations are run.
  In Set 1 (modelled roughness) $\alpha k = 1/40$ and
  in Set 2 (explicit sinusoidal roughness)
  $\lambda_x/k = \lambda_y/k \approx 7.1$.
  The grid spacing is uniform in the
  streamwise and spanwise direction, while the grid spacing in the
  wall-normal direction is stretched, with the finer wall grid spacing $\Delta z_w$
  and the coarser centreline grid spacing $\Delta z_c$.}
  \label{tab:cases}
  \end{center}
\end{table}

In the Set-1 simulations (modelled roughness),
the equations of motion, (\ref{eq:ns}), are numerically solved between
two no-slip, impermeable walls at $z = 0$, $2h$ and the effect of roughness is represented
by the parametric-forcing model of \cite{Busse+Sandham.2012}, whereby
a forcing term is added to the right-hand side of (\ref{eq:ns}\mbox{\textit{a}}) of the form,
\begin{equation}\label{eq:bodyforce}
  - \alpha F(z,k) u |u| \delta_{i1}, \quad \textrm{where} \quad
F(z,k) =
\left\{
    \begin{array}{ll}
      1, & z < k \quad \textrm{or} \quad 2h - k < z \\[2pt]
      0, & \textrm{otherwise}
    \end{array} \right.
\end{equation}
In principle, the effect of any roughness geometry, which includes
both pressure and viscous drag, can be formally written
as a body force on the right-hand side of the Navier--Stokes equation,
but for the present purposes, the form (\ref{eq:bodyforce}) is meant
to represent a hypothetical homogeneous roughness.
The roughness forcing is only active in the streamwise direction,
opposes the mean flow and always dissipates kinetic energy.
The roughness factor, $\alpha$, is thought to scale with the roughness density,
that is, the frontal area per unit volume
\citep{Nikora+coauthors.2007,Busse+Sandham.2012},
measured in inverse-length (area per unit volume) units.
Presently, $k = h/40$ and $\alpha = 1/(40k)$ for all simulations in Set 1
(refer to table \ref{tab:cases} for details).
A cosine mapping is used to stretch the grid in the wall-normal
direction.

\begin{figure}
\centerline{\includegraphics[width=\textwidth]{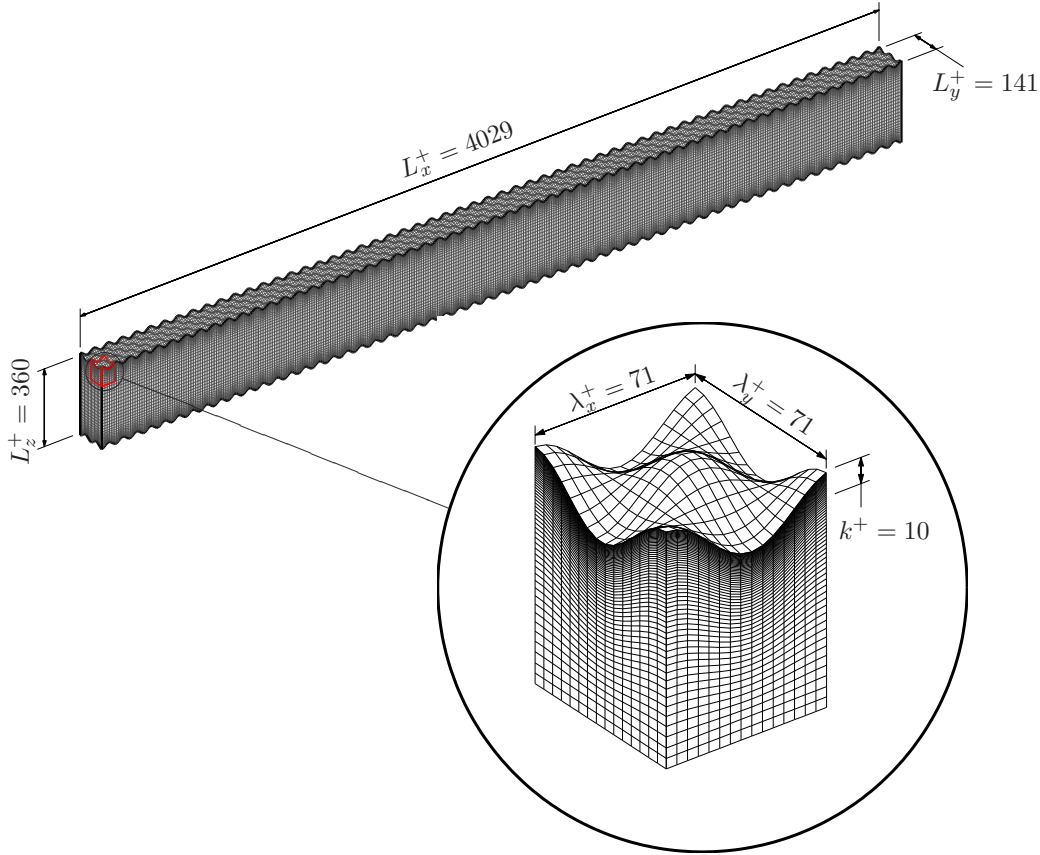}}
\caption{Grid for the Set-2 $\Rey_\tau \approx 180$ minimal-span simulation.}
\label{fig:explicitgrid}
\end{figure}
In the Set-2 simulations,
the sinusoidal roughness previously studied by \cite{Chan+coauthors.2015}
is explicitly represented by a roughness-conforming grid composed of hexahedral cells
(figure \ref{fig:explicitgrid}).
The grid is stretched in the wall-normal direction, with
the stretching ratio held below $1.05$.
Grid information for all cases in Set 2
is detailed in table \ref{tab:cases}.
The roughness surface described by the sinusoidal function,
\begin{equation}\label{eq:sinroughness}
z_w = k\cos(2\upi\,x/\lambda_x)\cos(2\upi\,y/\lambda_y),
\end{equation}
locates the no-slip, impermeable wall relative to
the reference plane at $z = 0$.
Under this definition, we consider
the characteristic roughness height $k$ to be the semi-amplitude of
this sinusoid.
For the present sinusoidal
roughness, the peak-to-valley roughness height $k_t$ ($= 2k$),
the root-mean-square roughness height $k_{rms}$ ($= k/2$) or other
physical roughness heights could instead be used.
Ultimately, the present method seeks to determine $k_s/k$ directly
and the specific choice of the physical roughness height
is therefore irrelevant,
as all one requires in practice is the ratio between the equivalent
sand-grain roughness and some readily
measurable geometric property of the rough surface.
For the present sinusoidal roughness, $k_s/k = 2k_s/k_{rms} = k_s/k_t/2$.
The roughness surface at the top wall is the mirror image
(across the centreline) of the roughness surface at the bottom wall.
Presently, $k = h/18$
and $\lambda_x = \lambda_y \approx 7.1k$
for all simulations in Set 2 (refer to table
\ref{tab:cases} for details).

\section{Results and discussion}
\subsection{Modelled roughness}\label{sec:modelledroughness}

\begin{figure}
    \centerline{\includegraphics[clip]{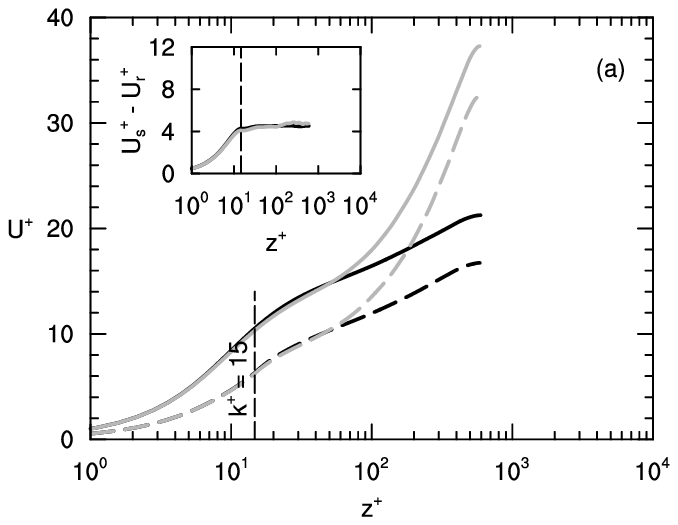}
               \includegraphics[clip]{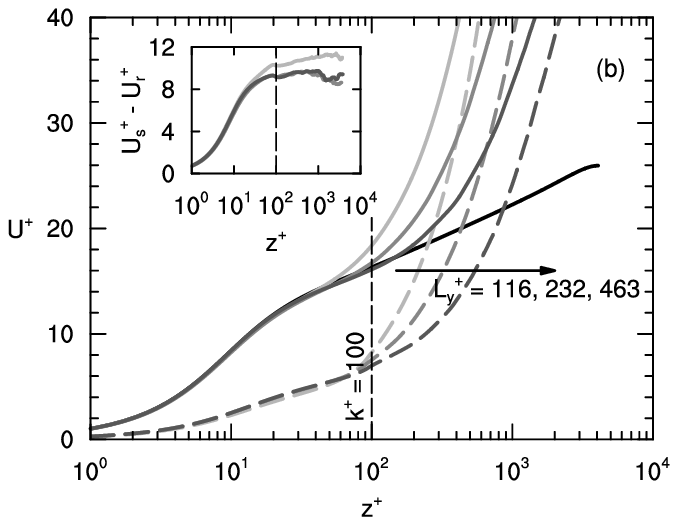}}
    \caption{Mean velocity profiles of simulated turbulent channel flow
    at (\mbox{\textit{a}}) $\Rey_\tau \approx 590$
    and (\mbox{\textit{b}}) $\Rey_\tau \approx 4000$: solid, smooth;
    dashed, modelled roughness; grey, minimal span; black, full span.
    The vertical dashed line marks the top of the roughness-forcing region.
    The inset shows that the velocity shift stays constant for both
    minimal- and full-span channels above the roughness-forcing region.
    The various grey profiles in (\mbox{\textit{b}}) correspond to the various minimal
    spans, $L_y^+ \approx 116$, $232$ and $463$. The inset in (\mbox{\textit{b}})
    shows that $L_y^+ \approx 116$
    is too narrow to accurately capture the roughness function for roughness
    heights above $k^+ = 100$.
    The full-span $\Rey_\tau \approx 4000 $ data in (\mbox{\textit{b}}) is from \cite{Bernardini+Pirozzoli+Orlandi.2014}.}
\label{fig:mvp}
\end{figure}
In figure \ref{fig:mvp}(\mbox{\textit{a}}), we show the mean velocity
profile from four simulations with modelled roughness (Set-1
simulations) at $\Rey_\tau \approx 590$,
corresponding to the smooth full-span, the rough full-span, the smooth
minimal-span and the rough minimal-span simulations.
Consistent with the findings of \cite{Hwang.2013}, figure \ref{fig:mvp}(\mbox{\textit{a}})
shows that the minimal span of $L_y^+ \approx 116$ is sufficient for
capturing the mean smooth-wall velocity profile given by the full-span simulations
up to the height $z_c^+ \approx 40$. We will use the notation $z_c^+$ to refer to the point above which the minimal profile departs from the full-span profile.
The present simulations show that this is also the case for the profile above
modelled rough walls (given by the dashed lines in figure \ref{fig:mvp}\mbox{\textit{a}}).
Above the roughness height of $k^+ = \Rey_\tau / (h/k)
\approx 590/40 \approx 15$, the inset of figure \ref{fig:mvp}(\mbox{\textit{a}})
shows that $\Delta U^+$ for the full-span channels is reproduced
by the minimal-span channels, even though $U^+$ itself above
$z_c^+ \approx 40$ for the minimal-span channels fails to capture
$U^+$ of the full-span channels.
We will use the notation $\Delta U^+$ to refer to the nominally constant value of
the downward velocity shift at matched $\Rey_\tau$, $U_s^+ - U_r^+$, evaluated above the roughness height.
The similarity of the mean relative velocity is, of course, a manifestation of Townsend's outer-layer similarity: the downwards shift in velocity is purely determined by the near-wall flow over the roughness, which sets the wall drag. This shift will remain the same irrespective of the state of the outer profile, which is vastly different for minimal- or full-span channels. These results demonsrate the efficacy
of the present method.
It is important to emphasise that the aforementioned
method to evaluate $\Delta U^+$ requires matched $\Rey_\tau$,
which is simple to achieve computationally since the outer
length scale and the driving pressure gradient (and therefore $U_\tau$)
can be fixed between smooth- and rough-wall simulations.
Experimentally, where $\Rey_\tau$ between smooth and rough walls
are seldom well-matched, $\Delta U^+$ is typically evaluated from
a shift in the log region (provided it exists).

\begin{figure}
    \centerline{\includegraphics[clip]{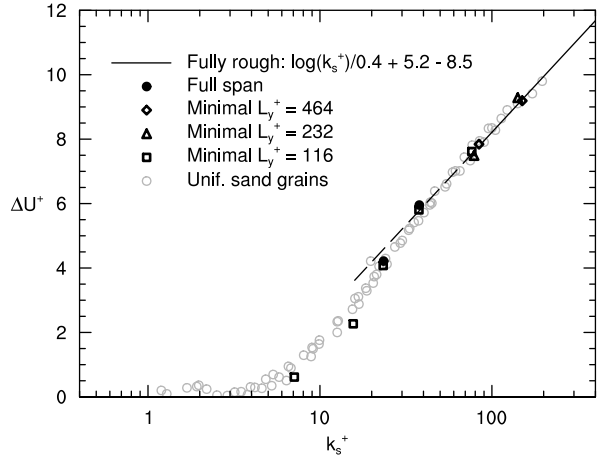}}
    \caption{The Hama roughness function at fixed $h/k = 40$
    for the modelled roughness, where $k_s/k \approx 1.6$, simulated
    in channels.
    The superimposed uniform sand-grain roughness data of \cite{Nikuradse.1933}
    are extracted from \cite{Jimenez.2004}.}
\label{fig:hama}
\end{figure}
When the top of the roughness-forcing region is higher than
$z_c^+$, a wider minimal channel is required.
A systematic study with various channel spans
at $\Rey_\tau \approx 4000$, as listed in table \ref{tab:cases},
studies this effect.
These results are shown in figure \ref{fig:mvp}(\mbox{\textit{b}}).
For the smooth walls, the data suggest that
the minimal channel is able to reproduce the full channel profiles
up to $z_c^+ \approx 0.4\,L_y^+$.
That is, the minimal profile from $L_y^+ \approx 116$ reproduces the
full profiles up to $z_c^+ \approx 0.4\,(116) \approx 46$,
the minimal profile from $L_y^+ \approx 232$ reproduces the full profiles
up to $z_c^+ \approx 0.4\,(232) \approx 93$,
the minimal profile from $L_y^+ \approx 463$ reproduces the full profiles
up to $z_c^+ \approx 0.4\,(463) \approx 185$, and the pattern
presumably goes on as long as $z_c^+ \approx 0.4\,L_y^+$ remains
in the log layer $z^+ <0.15\,\Rey_\tau$, where the size of eddies
is thought to scale with distance from the wall.
This behaviour is first reported by \cite{Flores+Jimenez.2010}, who suggested $z_c^+ \approx 0.3\,L_y^+$.
 We refer to the region $z < z_c$ as being `unconfined' since the turbulence in this region is faithfully captured, showing no signs of being constrained by the minimal span.
A difficulty arises when the top of the roughness rises above
the unconfined region (when $k > z_c$).
For the narrowest channel, we observe that the roughness
function is over-predicted, giving $\Delta U^+ \approx 11$
instead of the $\Delta U^+ \approx 9$ obtained for the two wider (yet still minimal)
channels.
This sensitivity leads to an additional constraint for the minimum
channel span, both of which must be satisfied
\begin{equation}
 L_y > k/0.4 \textrm{~~~~~and~~~~~} L_y^+ > 100
\end{equation}
Physically, this means that the channel must be wide
enough to fully immerse the roughness elements ($z_c > k$) in
natural, unconfined wall turbulence (which only occurs when $L_y^+ > 100$).
Although the constraint that $L_y > k/0.4$
is developed from the present modelled homogeneous roughness, where
$k$ is the modelled roughness height, we  
expect that, in general, $L_y > O(k)/0.4$ for any physical roughness height
$k$ such as $k_{rms}$ or $k_a$. In practice, a convergence study with
wider (still minimal) channels can be performed (cf.\
figure \ref{fig:mvp}\,\mbox{\textit{b}}) to manage this effect.
A conservative approach is to take $L_y > k_t/0.4$, where
$k_t$ is the maximum peak-to-valley roughness height.
The third constraint, requiring that $L_y > \lambda_y$ (where $\lambda_y$ is some characteristic spanwise length scale of the roughness texture, or repeating unit of the roughness) is not active for this case of modelled roughness, where the homogeneous forcing suggests $\lambda_y \rightarrow 0$.

A sweep in roughness Reynolds numbers, from
$k^+ \approx 5$ to $100$, corresponding to
$\Rey_\tau \approx 180$ to $4000$ (table \ref{tab:cases})
is needed in order to fully
characterise the roughness transition from the hydraulically
smooth to the fully rough regime. This in turn enables determination of the equivalent sand-grain roughness $k_s/k = C$, where $C$ is a constant that best scales the highest roughness data onto the fully rough asymptote. Through the use of minimal-span channels, even the simulations at $\Rey_\tau \approx 4000$,
are feasible.
The Hama roughness function for this sweep is presented in
figure \ref{fig:hama}. The data from the full-span simulations at $\Rey_\tau \approx 590$ and $950$ are shown by the filled circles. It is noted that the roughness function profile obtained from the minimal-span simulations matches very well with the full-span simulations.
Fitting to the fully rough regime, shown by the solid curve on figure \ref{fig:hama}, the present data show that
$k_s/k \approx 1.6$ for $\alpha = 1/(40k)$ and so we have characterised the
present modelled roughness from the transitionally rough regime,
$k_s^+ \approx 7$, to the fully rough regime, $k_s^+ \approx 160$. The obtained value for $k_s$ could now, in theory, be used to predict full-scale performance under this roughness condition \citep{G58,S2007}. Comparing the minimal data with the uniform sand-grain roughness data of \cite{Nikuradse.1933} shown in light grey markers, reveals that the parametric-forcing model 
with the box-profile shape function closely mimics the
roughness transition of uniform-sand-grain roughness {\citep[as previously noted by][]{Busse+Sandham.2012}}.

\subsection{Sinusoidal roughness}\label{sec:griddedroughness}
\begin{figure}
\centerline{\includegraphics{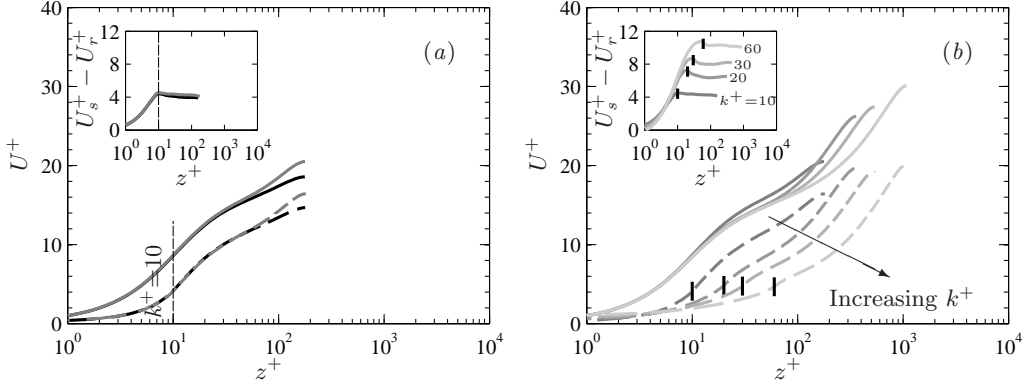}}
    \caption{Mean velocity profiles of turbulent channel flow:
    (\mbox{\textit{a}}) $\Rey_\tau \approx 180$, smooth and
    explicit roughness ($h/k = 18$) in
    minimal- and full-span channels; (\mbox{\textit{b}})
    $\Rey_\tau \approx 180$, $360$, $540$, $1080$, smooth and sinusoidal roughness
    ($h/k = 18$) in minimal-span channels.
    The vertical dashed line marks the top of the roughness element.
    The inset shows that the velocity shift stays the same for both
    minimal- and full-span channels above the roughness elements.}
\label{fig:mvp_grid_chan}
\end{figure}
Figure \ref{fig:mvp_grid_chan}(\mbox{\textit{a}}) is a plot of the mean
velocity profile from both the full and the minimal
$L_y^+ \approx 141$ channel with
explicitly represented sinusoidal roughness elements (Set-2 simulations)
at $\Rey_\tau \approx 180$.
Consistent with the simulations with modelled roughness, the mean profiles
from the minimal channel collapse onto the mean profiles from the full
channel below about $z^+_c \approx 0.4\,L_y^+ \approx 56$.
For $z^+ > 56$, the minimal profiles diverge from the
full channel, but the inset in \ref{fig:mvp_grid_chan}(\mbox{\textit{a}})
shows that the roughness function, $\Delta U^+$,
remains the same for both full and minimal channels.
In other words, the minimal channel is sufficient for predicting
the roughness function.
The method works in this case because
the top of the roughness elements
at $z^+ = k^+ \approx 10$ is well within the unconfined region $z^+ \approx 0.4\,L_y^+ \approx 56$,
and the roughness wavelength $\lambda_y^+$ $(\approx 71)$ is smaller than $L_y^+$.
An issue that often arises is the appropriate wall-normal location
to evaluate the roughness function.
In high-Reynolds-number flows, this location is frequently
taken to be somewhere in the log layer.
Presently, we observe that $U_s^+ - U_r^+$
is relatively insensitive to this location as long as $\Delta U^+$ is evaluated above the roughness height.
The insets in figures \ref{fig:mvp} and \ref{fig:mvp_grid_chan} demonstrate that this observation is consistent for all our results, for both explicit
and modelled roughness.

Figure \ref{fig:mvp_grid_chan}(\mbox{\textit{b}})
shows the mean velocity profiles for the minimal channel at a fixed blockage ratio $h/k = 18$, where $k^+$
is increased from $10$ to $60$ by increasing the friction Reynolds number from
$\Rey_\tau \approx 180$ to $1080$.
For the lowest $k^+ \approx 10$, the span of the minimal
channel is chosen to be $L_y^+ \approx 141$ in order to properly capture
the near-wall cycle and accommodate two periods of the sinusoidal
roughness.
In the cases with the higher roughness Reynolds numbers,
$k^+ \approx 20$, $30$ and $60$, the minimal channel
only captures one period of sinusoidal roughness, that is,
$\lambda_y = L_y$.
The corresponding predicted roughness functions $\Delta U^+$ are shown
in the inset of figure \ref{fig:mvp_grid_chan}(\mbox{\textit{b}}).

\begin{figure}
    \centerline{\includegraphics{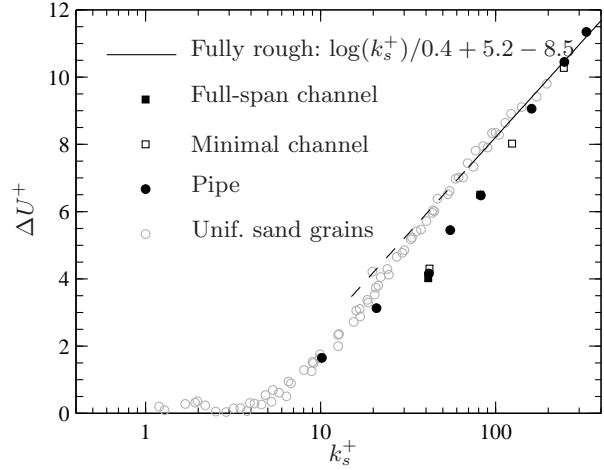}}
    \caption{The Hama roughness function at fixed $h/k = 18$
    for the sinusoidal roughness, where $k_s/k \approx 4.1$, simulated in
    channels and pipes.
    The pipe data is from \cite{Chan+coauthors.2015} and has the maximum blockage
    ratio, $R/k = 6.75$.
    The superimposed uniform sand-grain roughness data of \cite{Nikuradse.1933}
    are extracted from \cite{Jimenez.2004}.
   }
\label{fig:hama_grid}
\end{figure}
In order to assess the efficacy of the present method for an
explicitly represented sinusoidal roughness, we plot the Hama roughness function
versus the equivalent sand-grain roughness $k_s^+$ in figure
\ref{fig:hama_grid}.
Fitting the data in the fully rough regime, we are able to
characterise this particular sinusoidal roughness with
$\lambda_x = \lambda_y \approx 7.1\,k$ to find that
$k_s/k \approx 4.1$.
The minimal-channel predictions, shown by the open markers, collapse onto previously
studied full pipe-flow simulations of the same roughness by \cite{Chan+coauthors.2015}, albeit at various pipe blockage ratio $R/k$, as shown by the filled circles on figure \ref{fig:hama_grid}. This collapse further validates the ability of the minimal channel simulations to recover the same estimate for $k_s$ as given by full-size simulations.

It is worth pointing out that recent DNSs have also been simulated
in the fully rough regime \citep{Busse+Sandham.2012,Yuan+Piomelli.2014}.
An important distinction here is that these previous cases
used domain sizes that are deemed sufficiently large to capture the
full range of turbulent motions.
In contrast, the present minimal simulations have a severely
confined outer flow with much lower intensities and a radically
altered wake profile.
The potential saving in computational cost for the minimal simulations
could be redeployed to explicitly represent the roughness (see the
present Set-2 simulations), without the need for roughness models.

\section{Conclusions}\label{sec:conclusions}

We have presented a novel, fast and direct method for characterising
the hydraulic resistance of any given surface roughness.
The way in which a particular roughness transitions from the
hydraulically smooth to the fully rough regime is, to the
first approximation, described by how the roughness geometry
interacts with the near-wall flow.
The method presented herein shows that this interaction, so far as the mean drag is concerned,
is accurately captured 
using minimal-span channels.
There are other roughness effects, such as the change in turbulent kinetic energy,
and it remains to be seen the how these effects are represented by the minimal-channel technique.
We have validated the present method
with a specific sinusoidal roughness and showed that the
method is able to fully characterise the hydraulic (drag) behaviour of roughness
not only in the transitionally rough regime, but also in the fully rough
regime, yielding an estimate of $k_s$ which closely matches that given from full-span simulations.

The dynamic drag characterisation of a rough surface
is encapsulated in the Hama roughness function, which we show
can be accurately determined using minimal-span channels.
The fact that $\Delta U^+$ is plotted against $k^+$ (or $k_s^+$)
in the literature and not against $\textit{Re}_\tau$ acknowledges that
$\Delta U^+$, to a large extent, depends only on
the roughness-affected near-wall flow.
The minimal-span channel is a method for simulating this near-wall
flow of thickness $O(k)$ without resolving the outer scale $h \gg k$,
thereby breaking the curse of the (outer) Reynolds number.
The savings in computational cost for the present method
are possible because capturing only the near-wall flow
requires far less grid points than capturing the full flow.
For example, the full-span DNS
of \cite{Bernardini+Pirozzoli+Orlandi.2014}
at $\Rey_\tau \approx 4000$ requires $N_x \times N_y \times N_z
 = 8192 \times 4096 \times 1024 = 3.4\times 10^{10}$ grid points.
In contrast, the minimal-span ($L_y^+ \approx 463$)
DNS at the same $\Rey_\tau$ (table \ref{tab:cases}),
which is sufficient to reach the fully rough regime, requires only
$N_x \times N_y \times N_z
 = 384\times 96 \times 1024 = 3.8\times 10^7$ grid points,
or a three-orders-of-magnitude reduction in number of grid points.
The actual wall-clock time will, of course, depend on the
code and parallelisation, as well as the machine.
As an indicative comparison using the same code and machine,
the full-span ($L_y/h = \upi$) $\Rey_\tau \approx 950$ simulation (table \ref{tab:cases})
requires $2600\,\textrm{CPU hours}/(h/U_\tau)$,
while the minimal-span ($L_y^+\approx 112$) $\Rey_\tau \approx 950$ simulation (table \ref{tab:cases})
requires only $42\,\textrm{CPU hours}/(h/U_\tau)$.
These figures amount to a two-orders-of-magnitude reduction in CPU hours.

The present method can be used to characterise the
drag characteristics of many surfaces very quickly.
Such a powerful tool enables the researcher to now focus
on how the geometry of roughness affects the turbulent flow instead of
focussing on setting up expensive and time-consuming experiments,
physical or numerical.
Indeed, the present method even enables researchers to
reassess the physics underlying the success and failures
of previously proposed geometrical factors.
Recall that the minimal-span method retains the
same benefit from a full-span direct numerical simulation in
that the friction velocity formed from both viscous and
pressure contributions is directly measured.

In some ways, the present idea is not entirely new.
In the large-eddy simulation subgrid-scale modelling methodology,
the geometry-dependent large eddies are directly simulated
whilst the 
subgrid small eddies are assumed to be universal and are modelled or understood
through the energy-cascade phenomenology of Richardson,
Taylor and Kolmogorov.
Presently, this reasoning is reversed for the roughness problem, in which
the geometry-dependent small eddies are directly simulated
whilst the universal `supergrid' large eddies are assumed to be universal and are modelled or understood
through the outer-layer similarity hypothesis of Townsend, a kind
of small-eddy simulation, a term coined by \cite{Jimenez.2003}.

We have also provided guidelines on selecting the minimal channel. To obtain an accurate prediction of $\Delta U^+$, the roughness element must be submerged within unconfined near-wall flow, which yields the first two constraints: $L_y > k/0.4$ and $L_y^+ > 100$ respectively. Potentially, if $k$ were very large, the first of these constraints could lead to increasingly large `minimal' channels. However, in practice it seems unlikely that these simulations would need to be conducted for $k_s^+ \gg 100$ in order to establish the fully rough regime, and hence $L_y$ defined by this constraint should remain manageable. More problematic is the constraint that $L_y > \lambda_y$. For heterogeneous or sparse rough surfaces, the spanwise length scale or repeating unit $\lambda_y$ could potentially be very large in terms of viscous scaling, presenting an obvious Achilles' heel for this proposed method, although this would be equally true for full span simulations. Even for surfaces which have a small $\lambda_y$, yet where the roughness is randomly arranged (such as a sanded or painted surface) the minimal span would need to be sufficiently large to contain a statistically
representative sample of the rough surface. Even so this $\lambda_y$ would in most cases be significantly less that the $\upi h$ box width often required of full-span simulations. Within these constraints, this method could potentially be extremely useful for all rough surfaces that exhibit homogeneity over a relatively small scale. Examples could include surface finishes obtained from machining, painting or spray coating. One could also potentially use minimal channels scaled in this way to investigate drag reducing rough surfaces such as riblets. Indeed other passive or active surfaces where the primary effect of the surface is an alteration of the near-wall structure, or near-wall slip, could also be investigated using minimal channels (e.g.\ super-hydrophobic surfaces, compliant walls, porosity, acoustic liners).

We can also point to certain possible improvements or extensions of this technique in the future.  One issue with the minimal channels is that the centreline velocity becomes very high (relative to $U_\tau$), which adds computational expense through required reductions in the time step to satisfy the Courant--Friedrichs--Lewy (CFL) condition. A possible solution here would be to add a body-forcing term at the channel centreline to appropriately manage the magnitude of this velocity. An alternative is to use a simple eddy-viscosity term that is active sufficiently far above the roughness elements to account for absent eddies that are wider than the minimal span, which would reduce the centreline velocity. It is also possible that the number of grid points could be further halved by using an open channel \citep{Scotti.2006}. From a different approach, it could be the case that wider minimal channels may be more efficient in terms of obtaining statistically converged statistics for given CPU hours (due to reductions in the `burstiness' associated with minimal channels where $L_y^+$ is close to 100). Finally, and perhaps most promising, we are keen to explore the capability of a single temporal simulation, starting from the highest $\Rey_\tau$, and reducing the bulk velocity to map $\Delta U^+$ versus $k^+$ (and hence obtain $k_s^+$) in a single numerical experimental sweep.

\medskip
This research was undertaken on the NCI National Facility in Canberra,
Australia, which is supported by the Australian Commonwealth Government and also on the Victorian Life Science Computational Institute (VLSCI). The authors would like to gratefully acknowledge the financial support of the Australian Research Council (ARC).

\bibliographystyle{jfm}

\bibliography{minchannel}

\begin{thebibliography}{26}
\expandafter\ifx\csname natexlab\endcsname\relax\def\natexlab#1{#1}\fi

\bibitem[Bernardini {\em et~al.\/}(2014)Bernardini, Pirozzoli \&
  Orlandi]{Bernardini+Pirozzoli+Orlandi.2014}
{\sc Bernardini, M., Pirozzoli, S. \& Orlandi, P.} 2014 Velocity statistics in
  turbulent channel flow up to $\textit{Re}_\tau = 4000$. {\em J.~Fluid
  Mech.\/} {\bf 742}, 171--191.

\bibitem[Busse \& Sandham(2012)]{Busse+Sandham.2012}
{\sc Busse, A. \& Sandham, N.~D.} 2012 Parametric forcing approach to
  rough-wall turbulent channel flow. {\em J.~Fluid Mech.\/} {\bf 712},
  169--202.

\bibitem[Chan {\em et~al.\/}(2015)Chan, MacDonald, Chung, Hutchins \&
  Ooi]{Chan+coauthors.2015}
{\sc Chan, L., MacDonald, M., Chung, D., Hutchins, N. \& Ooi, A.} 2015 A
  systematic investigation of roughness height and wavelength in turbulent pipe
  flow in the transitionally rough regime. {\em J.~Fluid Mech.\/} p. In press.

\bibitem[Chin {\em et~al.\/}(2010)Chin, Ooi, Marusic \&
  Blackburn]{Chin+coauthors.2010}
{\sc Chin, C., Ooi, A. S.~H., Marusic, I. \& Blackburn, H.~M.} 2010 The
  influence of pipe length on turbulence statistics computed from direct
  numerical simulation data. {\em Phys.~Fluids\/} {\bf 22}, 115107.

\bibitem[Chung {\em et~al.\/}(2014)Chung, Monty \& Ooi]{Chung+Monty+Ooi.2014}
{\sc Chung, D., Monty, J.~P. \& Ooi, A.} 2014 An idealised assessment of
  townsend's outer-layer similarity hypothesis for wall turbulence. {\em
  J.~Fluid Mech.\/} {\bf 742}, R3.

\bibitem[Flack \& Schultz(2010)]{Flack+Schultz.2010}
{\sc Flack, K.~A. \& Schultz, M.~P.} 2010 Review of hydraulic roughness scales
  in the fully rough regime. {\em ASME J.~Fluids Eng.\/} {\bf 132}, 041203.

\bibitem[Flack {\em et~al.\/}(2005)Flack, Schultz \&
  Shapiro]{Flack+Schultz+Shapiro.2005}
{\sc Flack, K.~A., Schultz, M.~P. \& Shapiro, T.~A.} 2005 Experimental support
  for townsend's reynolds number similarity hypothesis on rough walls. {\em
  Phys.~Fluids\/} {\bf 17}, 035102.

\bibitem[Flores \& Jim\'enez(2010)]{Flores+Jimenez.2010}
{\sc Flores, O. \& Jim\'enez, J.} 2010 Hierarchy of minimal flow units in the
  logarithmic layer. {\em Phys.~Fluids\/} {\bf 22}, 071704.

\bibitem[Granville(1958)]{G58}
{\sc Granville, P.~S.} 1958 The frictional resistance and turbulent boundary
  layer of rough surfaces. {\em Tech. Rep.\/} 1024. Navy Department.

\bibitem[Hwang(2013)]{Hwang.2013}
{\sc Hwang, Y.} 2013 Near-wall turbulent fluctuations in the absence of wide
  outer motions. {\em J.~Fluid Mech.\/} {\bf 723}, 264--288.

\bibitem[Jim\'enez(2003)]{Jimenez.2003}
{\sc Jim\'enez, J.} 2003 Computing high-{Reynolds}-number turbulence: will
  simulations ever replace experiments? {\em J. Turbul.\/} {\bf 4}, 22.

\bibitem[Jim\'enez(2004)]{Jimenez.2004}
{\sc Jim\'enez, J.} 2004 Turbulent flows over rough walls. {\em
  Annu.~Rev.~Fluid Mech.\/} {\bf 36}, 173--196.

\bibitem[Jim\'enez \& Moin(1991)]{Jimenez+Moin.1991}
{\sc Jim\'enez, J. \& Moin, P.} 1991 The minimal flow unit in near-wall
  turbulence. {\em J.~Fluid Mech.\/} {\bf 225}, 213--240.

\bibitem[Lozano-Dur\'an \& Jim\'enez(2014)]{Lozano-Duran+Jimenez.2014}
{\sc Lozano-Dur\'an, A. \& Jim\'enez, J.} 2014 Effect of the computational
  domain on direct simulations of turbulent channels up to $\textit{Re}_\tau =
  4200$. {\em Phys.~Fluids\/} {\bf 26}, 011702.

\bibitem[Mahesh {\em et~al.\/}(2004)Mahesh, Constantinescu \&
  Moin]{Mahesh+Constantinescu+Moin.2004}
{\sc Mahesh, K., Constantinescu, G. \& Moin, P.} 2004 A numerical method for
  large-eddy simulation in complex geometries. {\em J.~Comput.~Phys.\/} {\bf
  197}, 215--240.

\bibitem[Moody(1944)]{M1944}
{\sc Moody, L.~F.} 1944 Friction factors for pipe flow. {\em Trans. ASME\/}
  {\bf 66}, 671--684.

\bibitem[Morinishi {\em et~al.\/}(1998)Morinishi, Lund, Vasilyev \&
  Moin]{Morinishi+coauthors.1998}
{\sc Morinishi, Y., Lund, T.~S., Vasilyev, O.~V. \& Moin, P.} 1998 Fully
  conservative higher order finite difference schemes for incompressible flow.
  {\em J.~Comput.~Phys.\/} {\bf 143}, 90--124.

\bibitem[Moser {\em et~al.\/}(1999)Moser, Kim \&
  Mansour]{Moser+Kim+Mansour.1999}
{\sc Moser, R.~D., Kim, J. \& Mansour, N.~N.} 1999 Direct numerical simulation
  of turbulent channel flow up to $\textit{Re}_\tau = 590$. {\em
  Phys.~Fluids\/} {\bf 11}, 943--945.

\bibitem[Nikora {\em et~al.\/}(2007)Nikora, McEwan, McLean, Coleman, Pokrajac
  \& Walters]{Nikora+coauthors.2007}
{\sc Nikora, V., McEwan, I., McLean, S., Coleman, S., Pokrajac, D. \& Walters,
  R.} 2007 Double-averaging concept for rough-bed open-channel and overland
  flows: theoretical background. {\em J.~Hydraul.~Eng.\/} {\bf 133}, 873--883.

\bibitem[Nikuradse(1933)]{Nikuradse.1933}
{\sc Nikuradse, J.} 1933 Laws of flow in rough pipes. {\em Tech. Rep.\/} 1292.
  NACA Tech.\ Mem.

\bibitem[Pope(2000)]{Pope.2000}
{\sc Pope, S.~B.} 2000 {\em Turbulent Flows\/}. Cambridge University Press.

\bibitem[Prandtl \& Schlichting(1955)]{PS55}
{\sc Prandtl, L. \& Schlichting, H.} 1955 The resistance law for rough plates.
  {\em Tech. Rep.\/} 258. Navy Department, translated by P. Granville.

\bibitem[Schultz(2007)]{S2007}
{\sc Schultz, M.~P.} 2007 Effects of coating roughness and biofouling on ship
  resistance and powering. {\em Biofouling\/} {\bf 23}, 331--341.

\bibitem[Scotti(2006)]{Scotti.2006}
{\sc Scotti, A.} 2006 Direct numerical simulation of turbulent channel flows
  with boundary roughened with virtual sandpaper. {\em Phys.~Fluids\/} {\bf
  18}, 031701.

\bibitem[Townsend(1976)]{Townsend.1976}
{\sc Townsend, A.~A.} 1976 {\em The Structure of Turbulent Shear Flow\/}, 2nd
  edn. Cambridge University Press.

\bibitem[Yuan \& Piomelli(2014)]{Yuan+Piomelli.2014}
{\sc Yuan, J. \& Piomelli, U.} 2014 Estimation and prediction of the roughness
  function on realistic surfaces. {\em J.~Turbul.\/} {\bf 15}, 350--365.

\end{thebibliography}

\end{document}